\documentclass[a4paper,showpacs,prb,twocolumn]{revtex4}%
\usepackage{amsfonts}
\usepackage{graphicx}
\usepackage{color}
\usepackage{amsmath}
\usepackage{amssymb}
\usepackage{latexsym}
\usepackage{psfrag}%
\begin{document}
\title{Suppression of compressible edge channels and spatial spin polarization in
the integer quantum Hall regime}
\author{S. Ihnatsenka}
\altaffiliation{Permanent address: Centre of Nanoelectronics,
Department of Microelectronics, Belarusian State University for
Informatics and Radioelectronics, 220013 Minsk, Belarus}
\affiliation{Solid State Electronics, Department of Science and
Technology (ITN), Link\"{o}ping University, 60174 Norrk\"{o}ping,
Sweden}
\author{I. V. Zozoulenko}
\affiliation{Solid State Electronics, Department of Science and Technology (ITN),
Link\"{o}ping University, 60174 Norrk\"{o}ping, Sweden}

\date{\today}

\begin{abstract}
We perform systematic numerical studies of the structure of
spin-resolved compressible strips in split-gate quantum wires
taking into account the exchange and correlation interactions
within the density functional theory in the local spin-density
approximation. We find that for realistic parameters of the wire
the exchange interaction can completely suppress the formation of
the compressible strips. As the depletion length or magnetic field
are increased, the compressible strips starts to form first for
the spin-down and then for spin-up edge channels. We demonstrate
that the widths of these strips plus the spatial separation
between them caused by the exchange interaction are equal to the
width of the compressible strip calculated in the Hartree
approximation for spinless electrons. We also discuss the effect
of electron density on the suppression of the compressible strips
in quantum wires.

\end{abstract}
 \maketitle

Quantitative analytical description of the edge states in terms of
compressible and incompressible strips given in seminal papers of Chklovskii
\textit{et al.}\cite{Chklovskii} has been the basis for understanding of
various features of the magnetotransport phenomena in the quantum Hall regime.
Purely electrostatic description of Chklovskii \textit{et al.}%
\cite{Chklovskii} does not however take into account the exchange
interaction between the electrons. This interactions is known to
dramatically affect the edge channel structure bringing about
qualitatively new features absent in a model of spinless
electrons. These features include, for example, spatial separation
of the edge states belonging to different spin
species\cite{Dempsey,Stoof}, and pronounced spin polarization of
the electron density in quantum wires exhibiting $1/B$-periodicity
related to the subband depopulation\cite{wire_polarization}. A
detailed knowledge of the structure of the compressible strips in
the presence of exchange interaction as well as information about
spatial separation of spin-up and spin-down edge channels are
necessary for the understanding and interpretation of a variety
magnetotransport phenomena such as tunnelling through quantum dot
and antidot structures in the edge state regime\cite{Andy,adot},
unusual periodicity of the Aharonov-Bohm oscillations in antidot
structures\cite{Ford} and many other\cite{Goldman}. The main
purpose of the present paper is to provide a quantitative
description of the structure of spin-resolved compressible strips
in split-gated quantum wires taking into account the exchange
interaction between the electrons. To this end we solve
numerically the Schr\"odinger equation with exchange and
correlation interactions included within the density functional
theory (DFT) in the local spin-density approximation
\cite{wire_polarization}. The choice of DFT for the description of
many-electron effects is motivated, on one hand, by its efficiency
in
practical implementation within a standard Kohn-Sham formalism\cite{Kohn_Sham}%
, and, on the other hand, by the excellent agreement between the
DFT and exact diagonalization\cite{Ferconi} and variational
Monte-Carlo calculations\cite{Rasanen,QDOverview} performed for
few-electron quantum dots. We find that for realistic parameters
of the wire the exchange interaction can completely suppress the
formation of the compressible strips. The exchange interaction
causes a spatial separation between the states of opposite spins,
with the separation distance being equal to the width of the
compressible strip computed in the Hartree approximation of
spinless electrons (which is well-described by Chklovskii
\textit{et al} theory\cite{Chklovskii}). As the depletion length
or magnetic field are increased, the compressible strips starts to
form first for the spin-down and then for spin-up states. We
demonstrate that in this case the widths of these compressible
strips plus the spatial spin separation between them due to the
exchange interaction are equal to the width of the compressible
strip for spinless electrons calculated in the Hartree
approximation.

We consider a quantum wire in a perpendicular magnetic field $B$ described by
the Hamiltonian $H=\sum_{\sigma}H^{\sigma},$
\begin{equation}
H^{\sigma}=H_{0}+V_{conf}(y)+V_{eff}^{\sigma}(y)+g\mu_{b}B\sigma,
\label{Hamiltonian}%
\end{equation}
where $H_{0}$ is the kinetic energy in the Landau gauge,
%$\mathbf{A}% =(-By,0,0)$,
%\begin{equation}
$H_{0}=-\frac{\hbar^{2}}{2m^{\ast}}\left\{  \left(
\frac{\partial}{\partial x}-\frac{eiBy}{\hbar}\right)
^{2}+\frac{\partial^{2}}{\partial y^2}\right\} ,\label{H_0} $
%\end{equation}
$\sigma=\pm\frac12 $ describes spin-up and spin-down states,
$\uparrow$ , $\downarrow$, and $m^{\ast}=0.067m_{e}$ is the GaAs
effective mass. The last term in Eq. (\ref{Hamiltonian}) accounts
for Zeeman energy where $\mu _{b}=e\hbar/2m_{e}$ is the Bohr
magneton, and the bulk $g$ factor of GaAs is $g=-0.44.$ In the
split-gate geometry the confining potential $V_{conf}(y)$ due to
the gates, donor layers and the Schottky barrier is well
approximated by the parabolic
confinement\cite{Stoof,wire_polarization},
$V_{conf}(y)=V_{0}+\frac{m^{\ast}}{2}\left(  \omega_{0} y\right)
^{2}$, where $V_{0}$ defines the bottom of the potential (we set
the Fermi energy $E_{F}=0).$
%Typical confinement strength for
%split-gate structures is in the range $\hbar\omega_{0}=1-5$
%meV\cite{Stoof,wire_polarization}.
The effective potential,
$V_{eff}^{\sigma}(y)$ within the framework of the Kohn-Sham
density functional theory reads \cite{Kohn_Sham},
\begin{equation}
V_{eff}^{\sigma}(\mathbf{r})=V_{H}(y)+V_{xc}^{\sigma}(y), \label{V_eff}%
\end{equation}
where $V_{H}($\textbf{r}$)$ is the Hartree potential due to the
electron density $n(y)=\sum_{\sigma}n^{\sigma}(y)$ (including the
mirror charges),\cite{wire_polarization}%
\begin{equation}
V_{H}(y)=-\frac{e^{2}}{4\pi\varepsilon_{0}\varepsilon_{r}}\int_{-\infty
}^{+\infty}dy^{\prime}n(y^{\prime})\ln\frac{\left(  y-y^{\prime}\right)  ^{2}%
}{\left(  y-y^{\prime}\right)  ^{2}+4b^{2}}, \label{V_H}%
\end{equation}
with $b$ being the distance from the electron gas to the interface
(we choose $b=60$ nm). For the exchange and correlation potential
$V_{xc}(y)$
%in the local spin density approximation is given by%
%\begin{equation}
%V_{xc}^{\sigma}=\frac{d}{dn^{\sigma}}\left\{
%n^{\sigma}\epsilon_{xc}\left(
%n,\zeta(y)\right)  \right\}  \label{LDA}%
%\end{equation}
%where
%$\zeta(y)=\frac{n^{\uparrow}-n^{\downarrow}}{n^{\uparrow}+n^{\downarrow
%}}$ is the local spin-polarization. In the present paper we
we utilize the widely used parameterization of Tanatar and
Cerperly\cite{TC} (see Ref.~[\onlinecite{wire_polarization}] for
explicit expressions for $V_{xc}^{\sigma}(y)$) . This
parameterization is valid for magnetic fields corresponding to the
filling factor $\nu>1$, which sets the limit of applicability of
our results. The spin-resolved electron density reads
\begin{equation}
n^{\sigma}(y)=-\frac{1}{\pi}\Im\int_{-\infty}^{\infty}dE\,G^{\sigma
}(y,y,E)f_{FD}(E-E_{F}), \label{density}%
\end{equation}
where $G^{\sigma}(y,y,E)$ is the retarded Green's function
corresponding to the Hamiltonian (\ref{Hamiltonian}) and
$f_{FD}(E-E_{F})$ is the Fermi-Dirac distribution function. The
Green's function of the wire, the electron and current densities
are calculated self-consistently using the technique described in
detail in Ref. [\onlinecite{wire_polarization}].

%=================================================================
\begin{figure}[ptb]
%[tb]
\includegraphics[scale = 0.9]{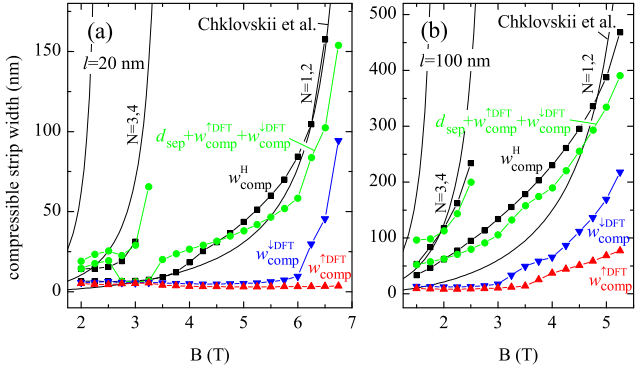}
\caption{(Color online) Width of the compressible strips in
quantum wires with different depletion lengths $l=20$ nm (a) and
100 nm (b). Parameters of the wires: $\hbar\omega_0$= 3 meV (a), 1
meV (b); $V_0=-0.2$ eV. The corresponding electron densities
$n(0)\approx 3\cdot 10^{15}\textrm{m}^{-2}$ and $2.5\cdot
10^{15}\textrm{m}^{-2}$.
Temperature $T=1$K. }%
\label{fig:width}%
\end{figure}
%=================================================================

\begin{figure*}[ptb]
%[tb]
\includegraphics[scale = 0.9]{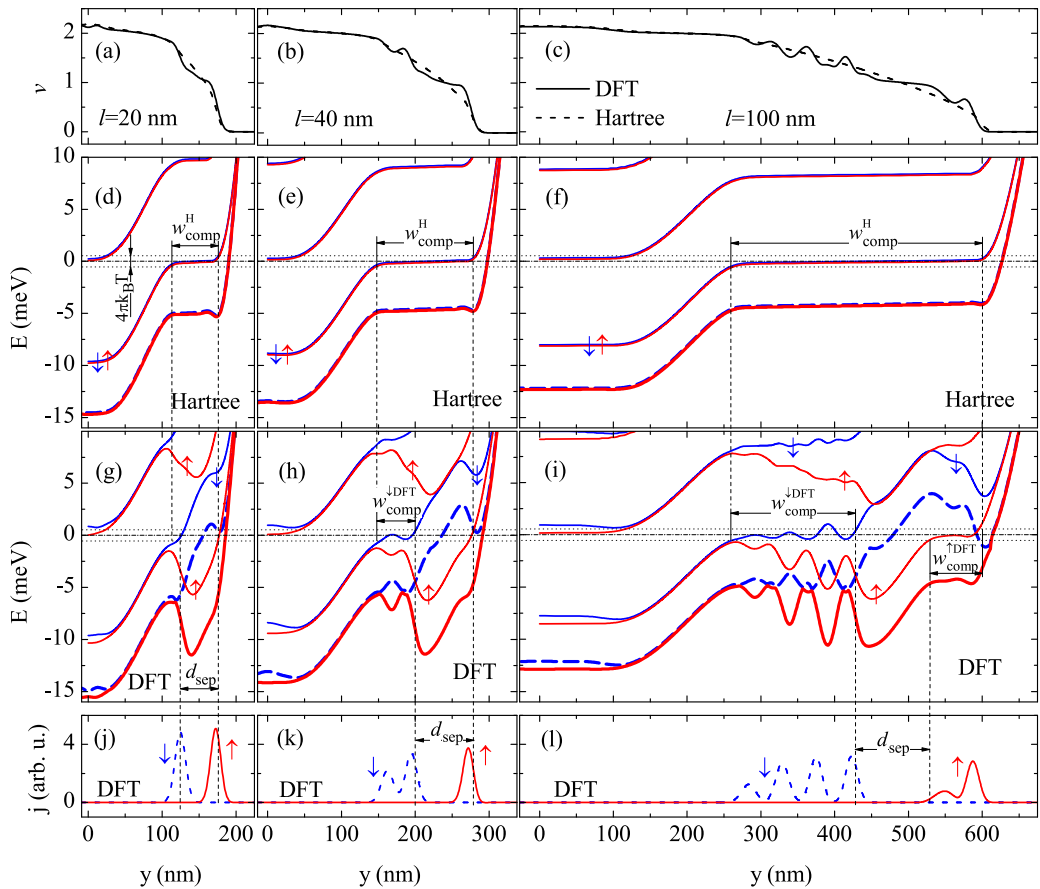}
\caption{ (Color online) (a)-(c) The electron density profile $\nu
(y)=n(y)/n_B$, (d)-(i)  the magnetosubband structure in Hartree
and DFT approximations and (j)-(l) the DFT current densities in
wires with different depletion lengths $l=$ 20 nm, 40 nm, 100 nm
(first, second and third columns respectively) for the filling
factor $\nu(0)\approx 2.2$. Fat solid and dashed lines show the
total confining potentials for spin-up and spin-down electrons
respectively. Parameters of the wires: $V_0=-0.2$ eV;
$\hbar\omega_0$= 3 meV, 2 meV, 1 meV (first, second and third
columns respectively). The corresponding magnetic fields are
$B\approx $5.9T, 5.4T, 4.9T; and the electron density $n(0)\approx
(3, 2.75, 2.5)\cdot 10^{15}\textrm{m}^{-2}$.  $T=1$K, $E_F=0$}%
\label{fig:band_diagram_l}%
\end{figure*}

\begin{figure}[ptb]
\includegraphics[scale = 0.8]{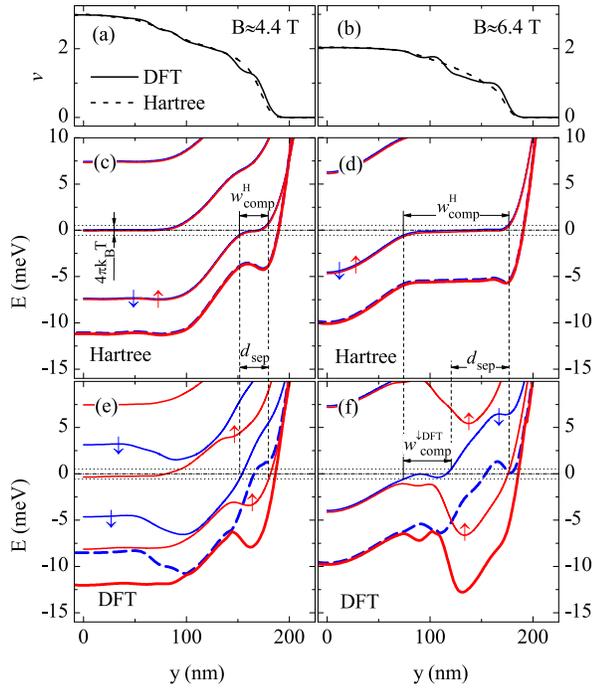}
\caption{(Color online) (a),(b) The electron density profile $\nu
(y)=n(y)/n_B$ and (c)-(f)  the magnetosubband structure in Hartree
and DFT approximations  in the wire with $l=20$nm for different
bulk filling factors $\nu (0)=3$ (left column) and $\nu (0)=2$
(right column) (corresponding to $B\approx 4.4$T and 6.4T
respectively). Fat solid and dashed lines show the total confining
potentials for spin-up and spin-down electrons respectively.
Parameters of the wire are $\hbar\omega_0=3\ \textrm{meV},\
V_{0}=-0.2\ \textrm{eV}$, corresponding to
$n(0)\approx 3\cdot 10^{15}\textrm{m}^{-2}$. $T=1$K, $E_F=0$.}%
\label{fig:band_diagram_nu}%
\end{figure}

\begin{figure}[tb]
\includegraphics[scale = 0.8]{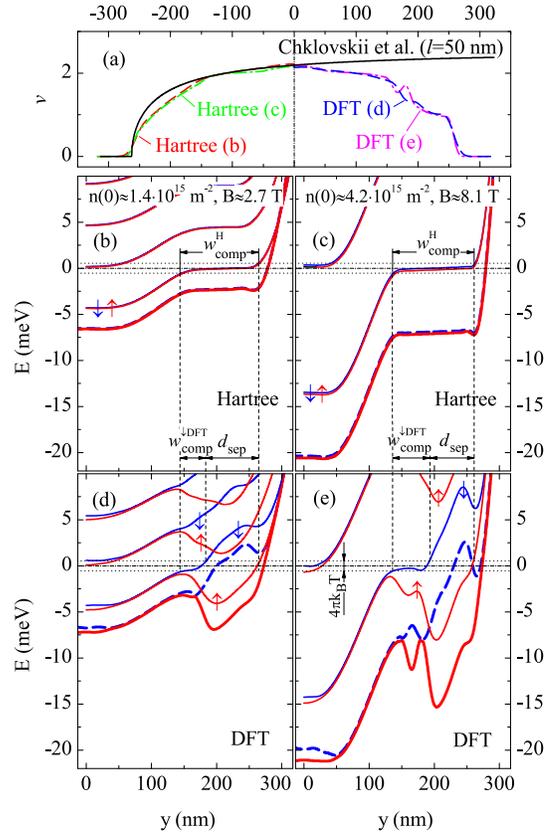}
\caption{(Color online) (a) The electron density profile $\nu
(y)=n(y)/n_B$, and (b)-(e) magnetosubband structure in Hartree and
DFT approximations for two \textit{different} quantum wires that
have the same depletion length $l=50$nm for the filling factor
$\nu (0)\approx 2.1$. Fat solid and dashed lines show the total
confining potentials for spin-up and spin-down electrons
respectively. Parameters of the wires, $\hbar\omega_0$ and
$V_{0}$, are (b),(d) 1.5 meV, -0.1 eV, and (c),(e) 2.6 meV, -0.3
eV. $T=1$K, $E_F=0$.}
\label{fig:same_l}%
\end{figure}

In order to outline the effect of the exchange interaction on the
structure of the compressible strips we first perform calculations
in the Hartree approximation (setting $V_{xc}^{\sigma}(y)=0$) and,
following Suzuki and Ando\cite{Ando_1998} compare them with the
predictions of Chklovskii \textit{et al.}\cite{Chklovskii}.
According to the Chklovskii \textit{et al.}
theory\cite{Chklovskii} the width and position of compressible and
incompressible strips are determined by the filling factor
$\nu(0)$ in the bulk (i.e in the middle of the wire, $y=0$), and
the depletion length $l$. Figure \ref{fig:width} shows the
magnetic field dependence of the width of the compressible strips,
$w_{comp}$, for two representative quantum wires of different
depletion lengths $l$. The corresponding electron density profiles
(local filling factors) $\nu(y)=n(y)/n_{B}$ $(n_{B}=eB/h)$ and
magnetosubband structures are illustrated in
Figs.~\ref{fig:band_diagram_l} and \ref{fig:band_diagram_nu} for
some representative values of the depletion lengths $l$ and
magnetic field respectively. (Following Suzuki and
Ando\cite{Ando_1998} we define the width of the compressible
strips $w_{comp}^{H}$ within the energy window $|E-E_{F}|<2\pi kT$
; the depletion length $l$ is extracted from the calculated
self-consistent density distribution by fitting to the Chklovskii
\textit{et al.}
dependence\cite{Chklovskii} $n(y)=n(0)\left(  \frac{y-l}%
{y+l}\right)  ^{1/2}$, see Fig.~\ref{fig:same_l}(a)). The
correspondence between the Chklovskii \textit{et al.} predictions
and Hartree calculations for $w_{comp}^{H}$ is good, being better
for wires with smaller depletion length $l$. This is related to
the fact that the parabolic confinement potential used in present
calculations is not fully equivalent to the split-gate Chklovskii
\textit{et al.} model \cite{Chklovskii}, especially for smooth
confinement (larger $l$). We finally note that in the considered
magnetic field interval the effect of Zeeman term on the subband
structure is negligible, such that we can refer to the Hartree
results as to the case of spinless electrons.

%=================================================================
%=================================================================

Having confirmed that the Hartree calculations are well-described
by Chklovskii \textit{et al.} model \cite{Chklovskii}, we turn to
the effect of the exchange interaction on the structure of the
compressible strips. Figure~\ref{fig:band_diagram_l} (a) shows the
electron density profiles %$\nu^{H(DFT)}(y)=n^{H(DFT)}(y)/n_{B}$
$\nu(y)=n(y)/n_{B}$ calculated in the Hartree  and DFT approaches
for a representative filling factor $\nu(0)\approx 2.2$. As
expected, the density profiles are practically the same, whereas
the corresponding subband structures are strikingly
different\cite{wire_polarization}. In what follows we shall
concentrate on the outermost spin-up and spin-down edge states
corresponding to the subbands $N=1,2$. (All the conclusions
reported below hold also for higher subbands). In the Hartree
approximation the spin-degenerate $N=1,2$ subbands form a
compressible strip of the width $w_{comp}^{H}$, see
Fig.~{\ref{fig:band_diagram_l}} (d).
Figure~{\ref{fig:band_diagram_l}} (g) shows corresponding subband
structure in the DFT approximation, where exchange interaction
lifts the spin degeneracy by pushing the spin-up and spin-down
subbands respectively below and above Fermi energy. This occurs
because the exchange potential for spin-up electrons depends on
the density of spin-down electrons and vice
versa\cite{TC,wire_polarization}. In the compressible region the
subbands are only partially filled (because $f_{FD}%
<1$ in the the window $|E-E_{F}|\lesssim2\pi kT$), and, therefore,
the population of spin-up and spin-down subbands can be different.
In the DFT calculation, this population difference (triggered by
Zeeman splitting) is strongly enhanced by the exchange interaction
leading to different effective potentials for spin-up and
spin-down electrons. This causes the separation of the subbands
which magnitude can be comparable to the Landau level spacing
$\hbar\omega$. As the result, the compressible region (present in
the Hartree approximation) is suppressed and the spin-up and
spin-down states become spatially separated by the distance
$d_{sep}\approx w_{comp}^{H}$. This is illustrated in
Fig.~\ref{fig:band_diagram_l} (j) which shows the current
densities for the outermost spin-up and spin-down channels, peaked
at the positions where the corresponding spin-up and spin-down
subbands intersect the Fermi energy. (The current densities in the
Hartree approximation are practically degenerated and delocalized
within the whole extension of the compressible strip). Outside
this region the subbands remains degenerate because they are fully
occupied when $E\lesssim E_{F}-2\pi kT$. As the result, the
corresponding spin-up and spin-down densities are the same, hence
the exchange and correlation potentials for spin-up and spin-down
electrons are equal, $V_{xc}^{\uparrow
}(y)=V_{xc}^{\downarrow}(y)$. (Note that towards the center of the
wire the degeneracy of $N=1,2$ subbands is lifted again because of
the difference of spin-up and spin-down densities for electrons in
the subbands $N=3,4$.)

Evolution of the subband structure in quantum wires as the
depletion length $l$ is increased is shown in
Fig.~\ref{fig:band_diagram_l}. The described above scenario of the
suppression of the compressible strips holds also for larger $l$
and $B$, with one new important feature. According to the
electrostatic description of Chklovskii \textit{et
al.}\cite{Chklovskii}, the compressible strips are more easily
formed in a structure with larger $l$, which is confirmed
experimentally\cite{Oto}. This feature is clearly manifested in
Hartree calculations, Figs.~\ref{fig:band_diagram_l} (d)-(f),
where the width of the compressible strip grows as $l$ is
increased (see also Fig.~\ref{fig:width}). The same effect holds
true in the presence of the
exchange interaction. When $l$ is small as in Figs.~\ref{fig:band_diagram_l}%
(g), the exchange interaction completely suppresses the
compressible strip and the spin-up and spin-down spin channels
become spatially separated with $d_{sep}\approx w^{H}_{comp}$ as
described above. For larger $l$ the compressible strip of the
width $w^{\downarrow DFT}_{comp}$ starts to form for the spin-down
subband (Fig.~\ref{fig:band_diagram_l} (h)), such that $d_{sep}+
w^{\downarrow DFT}_{comp}\approx w^{H}_{comp}$, see
Fig.~\ref{fig:band_diagram_l}, second column. With further
increase of the depletion length $l$ the compressible strip of the
width $w^{\uparrow DFT}_{comp}$ starts to form also for the
spin-up edge channel, see Fig.~\ref{fig:band_diagram_l}, third
column). Note that decrease of the spatial separation between the
edge channels by steeping the confining potential was
experimentally demonstrated by M\"{u}ller \textit{et
al}\cite{Muller}.

%=================================================================
%=================================================================

Qualitatively similar evolution of the magnetosubbands takes place
when the magnetic field is increased. This is illustrated in
Fig.~\ref{fig:band_diagram_nu} for two representative filling
factors $\nu(0)$, where the exchange interaction completely
suppress the compressible strip for $\nu(0)=3$, whereas for
$\nu(0)=2$ the compressible strip starts to form for the spin-down
channel. The effect of the exchange interaction on the structure
of the compressible strips is summarized in Fig.~\ref{fig:width}:
\textit{the spatial separation between spin-up and spin down edge
channels plus the widths of the corresponding compressible strips
are equal to the width of the compressible strip calculated in the
Hartree approximation}, $d_{sep}+ w^{\uparrow
DFT}_{comp}+w^{\downarrow DFT}_{comp} \approx w^{H}_{comp}$.
%(We stress again that is $w^{H}_{comp}$ is
%well-described by Chklovskii \textit{et al.} electrostatic
%picture\cite{Chklovskii}, $w^{H}_{comp}\approx w^{Chk}_{comp}$.)

It should be also noted that relatively weak spatial spin
polarization due to the exchange interaction occurs even for the
fields when the width of the compressible strip is negligible,
$w^{H}_{comp}\approx0$. This can be seen in Fig.~\ref{fig:width}
(a) for $B\lesssim4$T.

%=================================================================
%=================================================================

Let us now discuss the effect of the electron density on
suppression of the compressible strips. According to Chklovskii
\textit{et al.} theory\cite{Chklovskii}, at the given filling
factor $\nu(0)$, the depletion length $l$ represents the only
relevant scale determining the electron density profile. This
means that widths of the compressible strips $w_{comp}^H$ in
different wires with different confinement strengths $\hbar\omega$
and electron densities $n(0)$ are expected to be the same as soon
as their depletion lengths are the same. On the other hand, the
exchange interaction is density dependent, favoring stronger spin
polarization for lower electron densities\cite{TC,Martin}. It is
expected therefore that the effect of the exchange interaction
will be reduced in wires with larger densities.
Figure~\ref{fig:same_l} shows the density profiles and
magnetosubband structure for two \textit{different} quantum wires
characterized by the same depletion length $l$. The corresponding
Hartree density profiles (as well as $w_{comp}^H$) are practically
the same, whereas the DFT density profiles show some difference.
In accordance to the expectation, the width of the compressible
strip $w_{comp}^{DFT}$ is larger for higher electron density, cf.
Figs. \ref{fig:same_l} (d) and (e).

To conclude, we perform a detailed numerical study of the effect
of the exchange interaction on the structure of compressible edge
channels, and we find that this interaction can completely or
partly suppress the formation of the compressible strips.

S. I. acknowledges financial support from the Royal Swedish Academy of
Sciences and the Swedish Institute.

\end{document}